\documentstyle[12pt,epsf,epsfig]{article}
\def\fo{\hbox{{1}\kern-.25em\hbox{l}}}

\def\beq{\begin{equation}}
\def\eeq{\end{equation}}
\def\eq{\end{equation}}
\def\to{\rightarrow}

\def\bsg{\ifmmode B\to X_s\gamma\else $B\to X_s\gamma$\fi}
\def\bsll{\ifmmode B\to X_s\ell^+\ell^-\else $B\to X_s\ell^+\ell^-$\fi}
\def\bstt{\ifmmode B\to X_s\tau^+\tau^-\else $B\to X_s\tau^+\tau^-$\fi}
\def\shat{\ifmmode \hat{s}\else $\hat{s}$\fi}

\newcommand{\newc}{\newcommand}

\newc{\lcal}{\int {\cal L}dt}

\newc{\LSP}{{\chi^0_1}}
\newc{\stauR}{{\tilde \tau_R}}
\newc{\stau}{{\tilde \tau_1}}
\newc{\mstop}{m_{\tilde{t}}}
\newc{\mHpm}{m_{H^\pm}}
\newc{\gsim}{\lower.7ex\hbox{$\;\stackrel{\textstyle>}{\sim}\;$}}
\newc{\lsim}{\lower.7ex\hbox{$\;\stackrel{\textstyle<}{\sim}\;$}}
\newc{\ie}{{\it i.e.}}
\newc{\etal}{{\it et al.}}
\newc{\eg}{{\it e.g.}}
\newc{\kev}{\hbox{\rm\,keV}}
\newc{\mev}{\hbox{\rm\,MeV}}
\newc{\gev}{\hbox{\rm\,GeV}}
\newc{\tev}{\hbox{\rm\,TeV}}
\newc{\xpb}{\hbox{\rm\, pb}}
\newc{\xfb}{\hbox{\rm\, fb}}

\newc{\mtop}{m_t}
\newc{\mbot}{m_b}
\newc{\mz}{m_Z}
\newc{\mw}{M_W}
\newc{\alphasmz}{\alpha_s(m_Z^2)}
\newc{\swsq}{\sin^2\theta_W}
\newc{\tw}{\tan\theta_W}
\newc{\cw}{\cos\theta_W}
\newc{\sw}{\sin\theta_W}
\newc{\BR}{\hbox{\rm BR}}
\newc{\zbb}{Z\to b\bar}
\newc{\Gb}{\Gamma (Z\to b\bar b)}
\newc{\Gh}{\Gamma (Z\to \hbox{\rm hadrons})}
\newc{\rbsm}{R_b^\hbox{\rm sm}}
\newc{\rbsusy}{R_b^\hbox{\rm susy}}
\newc{\drb}{\delta R_b}

\newc{\sgn}{\mbox{sgn}}
\newc{\tbeta}{\tan\beta}
\newc{\uL}{{\tilde u_L}}
\newc{\uR}{{\tilde u_R}}
\newc{\cL}{{\tilde c_L}}
\newc{\cR}{{\tilde c_R}}
\newc{\tL}{{\tilde t_L}}
\newc{\tR}{{\tilde t_R}}
\newc{\dL}{{\tilde d_L}}
\newc{\dR}{{\tilde d_R}}
\newc{\sL}{{\tilde s_L}}
\newc{\sR}{{\tilde s_R}}
\newc{\bL}{{\tilde b_L}}
\newc{\bR}{{\tilde b_R}}
\newc{\eL}{{\tilde e_L}}
\newc{\eR}{{\tilde e_R}}
\newc{\mhp}{m_{H^\pm}}
\newc{\mhalf}{m_{1/2}}
\newc{\emt}{{e/\mu /\tau}}

\newc{\lR}{\tilde{l}_R}
\newc{\lL}{\tilde{l}_L}
\newc{\nL}{\tilde{\nu}_L}
\newc{\na}{\chi^0_1}
\newc{\nb}{\chi^0_2}
\newc{\nc}{\chi^0_3}
\newc{\nd}{\chi^0_4}
\newc{\ca}{\chi^{\pm}_1}
\newc{\cb}{\chi^{\pm}_2}
\newc{\camp}{\chi^\mp_1}
\newc{\cbmp}{\chi^\mp_1}
\newc{\capos}{\chi^{+}_1}
\newc{\caneg}{\chi^{-}_1}
\newc{\phit}{\phi_t}
\newc{\phib}{\phi_b}
\newc{\phiew}{\phi_{ew}}
\newc{\htz}{h^0_t}
\newc{\hbz}{h^0_b}
\newc{\hewz}{h^0_{ew}}
\newc{\hsmz}{h^0_{sm}}
\newc{\huz}{h^0_u}
\newc{\hsusyz}{h^0_{susy}}

\def\mp{M_P}

\def\beq{\begin{equation}}
\def\eeq{\end{equation}}
\def\bea{\begin{eqnarray}}
\def\eea{\end{eqnarray}}
\def\slashchar#1{\setbox0=\hbox{$#1$}           
   \dimen0=\wd0                                 
   \setbox1=\hbox{/} \dimen1=\wd1               
   \ifdim\dimen0>\dimen1                        
      \rlap{\hbox to \dimen0{\hfil/\hfil}}      
      #1                                        
   \else                                        
      \rlap{\hbox to \dimen1{\hfil$#1$\hfil}}   
      /                                         
   \fi}                                         %
\catcode`@=11
\long\def\@caption#1[#2]#3{\par\addcontentsline{\csname
  ext@#1\endcsname}{#1}{\protect\numberline{\csname
  the#1\endcsname}{\ignorespaces #2}}\begingroup
    \small
    \@parboxrestore
    \@makecaption{\csname fnum@#1\endcsname}{\ignorespaces #3}\par
  \endgroup}
\catcode`@=12

\newcommand{\tinySUSY}{\makebox[0.85cm][l]{$\line(4,1){19}$\hspace{-0.77cm}
{\tiny{SUSY}}}}

\def\simlt{\stackrel{<}{{}_\sim}}
\def\simgt{\stackrel{>}{{}_\sim}}

\begin{document}

\baselineskip=18pt

\begin{titlepage}
\begin{flushright}
IZTECH-P/2006-01\\
IPM-2006-004
\end{flushright}

\begin{center}
\vspace{1cm}

{\Large \bf Correlating  $\mu$ Parameter and Right-Handed Neutrino Masses in  $N=1$ Supergravity}

\vspace{0.5cm}

{ D. A. Demir$^{1}$\footnote{\baselineskip=16pt E-mail address:
{\tt demir@physics.iztech.edu.tr}}} and { Y.
Farzan$^{2}$\footnote{\baselineskip=16pt E-mail address: {\tt
yasaman@theory.ipm.ac.ir}}
\hspace{3cm}\\
 $^{1}$~{\small Department of Physics, Izmir Institute of Technology, TR35430, Izmir, Turkey}.
\hspace{0.3cm}\\
 $^{2}$~{\small Institute for Studies in Theoretical Physics and Mathematics (IPM),
P.O. Box 19395-5531, \\ Tehran, Iran}.}

\end{center}
\vspace{1cm}

\begin{abstract}
\medskip
The minimal supersymmetric standard model, when extended to embed
the seesaw mechanism, obtains two dimensionful parameters in its
superpotential: the $\mu$ parameter and the right-handed neutrino
mass $M_N$. These mass parameters, belonging to the supersymmetric
sector of the theory, pose serious naturalness problems as their
scales are left completely undetermined. In fact, for correct
phenomenology, $\mu$ must be stabilized at the electroweak scale
while $M_N$ lies at an intermediate scale. In this work we
construct an explicit model of the hidden sector of $N=1$
supergravity for inducing both $\mu$ and $M_N$ at their right
scales. The model we build utilizes lepton number conservation and
continuous $R$ invariance as two fundamental global symmetries to
forbid bare $\mu$ and $M_N$ appearing in the superpotential, and
induces them at phenomenologically desired scales via spontaneous
breakdown of the global symmetries and the supergravity. We
discuss briefly various phenomenological implications of the
model.
\end{abstract}

\bigskip
\bigskip

\begin{flushleft}
IZTECH-P/2006-01 \\
IPM 2006-004\\
January 2006
\end{flushleft}

\end{titlepage}

\tableofcontents

\section{Introduction}

Supergravity, once spontaneously broken at a scale
$M_{\tinySUSY}$~, gives rise to a softly-broken globally
supersymmetric theory at a scale $M_{\tinySUSY}^2/M_{Pl}$ which
corresponds to the mass scale of the gravitino, $m_{3/2}$
\cite{sugra} (notice that $M_{\tinySUSY}$ differs from soft masses
often denoted by $m_{SUSY}$) . If the gravitino mass is  at the
weak scale, $m_{3/2}\sim (1-10)M_{EW}$, or equivalently, if
supergravity is spontaneously broken at an intermediate scale
$M_{\tinySUSY}^2\sim m_{3/2} M_{Pl}$, the gauge hierarchy problem
is solved: The fact that supersymmetry is broken only softly
guarantees that the electroweak scale is radiatively stable, that
is, the ratio $M_{EW}/M_{Pl}\sim m_{3/2}/M_{Pl}$ is immunized
against quantum fluctuations.

The most economic description of the observable sector is realized
by the Minimal Supersymmetric Standard  Model (MSSM) which
essentially corresponds to a direct supersymmetrization of the SM
spectrum. One of the crucial aspects of the entire mechanism is
the breakdown of local supersymmetry at an intermediate scale
$M_{\tinySUSY}^2\sim m_{3/2} M_{Pl}$ which itself poses a new
naturalness problem. The question of how such an intermediate
scale has been formed can be answered only through a concrete
modeling of the hidden sector. Just to give an idea, one can
consider, for instance, dynamical supersymmetry breaking scenarios
in which all energy scales in the infrared are generated from
$M_{Pl}$ via dimensional transmutation \cite{dynamic}. Right here
one recalls that intermediate scales like $M_{\tinySUSY}$ are
 also necessitated by other phenomena not related to supersymmetry
breaking. As an example, one can allude to the Peccei-Quinn
mechanism \cite{pq} which is devised to solve  the strong CP
problem. This mechanism is based on the presence of  an
intermediate scale $M_{PQ} \sim 10^{14}\ {\rm GeV}$.

Another example, on which we are going to concentrate  in this
paper, is the famous seesaw mechanism \cite{seesaw} which explains
the tiny but nonzero neutrino masses. The seesaw mechanism is
based on the existence of ultra heavy right-handed neutrinos,
$N_i$, which are singlets of the SU(3)$\times$SU(2)$\times$U(1)
symmetry of the standard model. Indeed, in spite of several
alternative models \cite{smirnov}, the seesaw mechanism is
arguably the most popular way of explaining the tiny masses of
neutrinos, partly because it provides an explanation for the
baryon asymmetry of the universe through a mechanism called
leptogenesis \cite{lepto}. Successful leptogenesis requires masses
of heavy right-handed neutrinos to be above $\sim 10^9$~GeV
\cite{buchmuller}.

Obviously, low-energy phenomena do not
necessitate any correlation among the mass scales $M_{\tinySUSY}$,
$M_{PQ}$ and $M_{N}$. They show up as independent scales, needed to
explain distinct phenomena. However, it would establish a strong
case, besides superstrings, for the existence of a supersymmetric
organizing principle operating at ultra high energies if they can
be correlated within a specific model. Concerning this point, one
here recalls \cite{kimnilles} in which $M_{PQ}$ and
$M_{\tinySUSY}$ have been correlated by using a hidden sector
composite axion. The subject matter of the present work will be
essentially  to relate $M_{\tinySUSY}$ and $M_{N}$, leaving aside
$M_{PQ}$, within $N=1$ supergravity.

The superpotential of MSSM-RN is given by
\begin{eqnarray}
\label{superpot} \widehat{W}_{MSSM-RN} &=& \mu \widehat{H}_u\cdot
\widehat{H}_d + M_{N} \widehat{N}^c \widehat{N}^c\nonumber\\ &+&
Y_u \widehat{Q}\cdot \widehat{H}_u \widehat{U}^c + Y_d
\widehat{H}_d\cdot \widehat{Q} \widehat{D}^c + Y_e
\widehat{H}_d\cdot \widehat{L} \widehat{E}^c + Y_{\nu}
\widehat{L}\cdot \widehat{H}_u \widehat{N}^c
\end{eqnarray}
where $\widehat{N}^c$ stands for the anti right-handed neutrino
supermultiplet \cite{seesaw}. This model classically preserves the
baryon number while $R$ parity and SM gauge group are exact
symmetries of the model even at the quantum level.

The superpotential (\ref{superpot}) involves two dimensionful
parameters $\mu$ and $M_N$. These mass parameters are both nested
in the superpotential, and thus, they bear no relation whatsoever
to the supersymmetry breaking sector of the theory. Therefore,
they pose a serious naturalness problem in that MSSM offers no
mechanism, dynamical or otherwise, to enable one to know
characteristic scales of $\mu$ and $M_N$. In fact, present
neutrino data already require $M_N \sim \langle H_u \rangle^2
Y_\nu^2/m_\nu \simgt 10^{14} (Y_\nu)^2\ {\rm GeV}$. On the other
hand, LEP lower bound on chargino mass \cite{lepsite} leads in a
rather model independent way to a lower bound on the $\mu$
parameter: $\mu > 110\ {\rm GeV}$ \cite{lower}. On the other
hand, to have a successful electroweak symmetry breaking $\mu$ has
to be stabilized at the weak scale \cite{muprob}. Consequently, the
MSSM must be extended to provide a dynamical understanding of how
and why $\mu$ and $M_N$ are stabilized to their phenomenologically
favored scales. This is the naturalness problem we discuss in this
work. Actually, part of the problem $i.e.$ stabilization of the
$\mu$ parameter at the ${\rm TeV}$ scale has already been
discussed in various contexts and various solutions have been
devised \cite{muprob}. What is left over is to understand the
mechanism which stabilizes $M_N$ at an intermediate scale of $\sim
10^{14}\ {\rm GeV}$. In what follows we will attack on both
naturalness problems by constructing a hidden sector model which
leads to a dynamical determination of $\mu$ and $M_N$ upon
breakdown of certain global symmetries \cite{dimopoulos} and
supergravity \cite{gm}. It will turn out to be a hybrid model in
that we will utilize both K{\"a}hler potential and superpotential
of the hidden sector fields such that induction of the $\mu$
parameter proceeds in a way similar to the Giudice-Masiero mechanism \cite{gm}.

In Sec. 2, we will attempt to formulate a model of hidden sector
dynamics which induces $\mu$ and $M_N$ {\it simultaneously}, as a
result of  spontaneous local supersymmetry breaking. As will be
seen, this can be accomplished by including logarithmic terms in
the K\"ahler potential (which might be motivated by string theory
\cite{string}). In Sec. 3, we will provide a brief discussion of
certain phenomenological implications of this model. In Sec. 4, we
summarize our conclusions.

\section{The Model}

Any attempt at answering the question put forward in the introduction
should first provide a way of forbidding bare
$\mu$ and $M_N$ appearing in the superpotential. The fact that $\widehat{W}_{MSSM-RN}$ possesses additional
continuous symmetries in the absence of $\mu$ and $M_N$ \cite{dimopoulos} implies that imposing global symmetries
 can
protect $\widehat{W}_{MSSM-RN}$ against the bare $\mu$ and $M_N$
parameters. In particular, imposing  a global continuous $R$
invariance and restoring lepton number conservation forbid the
bare $\mu$ and $M_N$ parameters, respectively.  This can be
achieved by introducing new fields appropriately charged under
these symmetries. Realization of this setup requires at least two
hidden sector chiral superfields $\widehat{z}_1$, $\widehat{z}_2$
with charges $R(\widehat{z}_1) = 0$, $R(\widehat{z}_2) = 2$,
$L(\widehat{z}_1)=2$ and $L(\widehat{z}_2) = -2$.  The $R$ charge
of each lepton and quark superfield equals $1$ and
$R(\widehat{H}_u) = R(\widehat{H}_d) = 0$ so that the
superpotential (\ref{superpot}) acquires two units of $R$ charge.
The complete superpotential can be decomposed as
\begin{eqnarray}
\widehat{W} = \widehat{W}_{obs} + \widehat{W}_{hid} + \widehat{W}_{obs-hid}
\end{eqnarray}
where
\begin{eqnarray}\label{yukawa}
\widehat{W}_{obs} &=& Y_u \widehat{Q}\cdot \widehat{H}_u
\widehat{U}^c + Y_d \widehat{H}_d\cdot \widehat{Q} \widehat{D}^c +
Y_e \widehat{H}_d\cdot \widehat{L} \widehat{E}^c + Y_{\nu}
\widehat{L}\cdot \widehat{H}_u
\widehat{N}^c\nonumber\\
\widehat{W}_{hid} &=& M_{hid} \widehat{z}_2 \widehat{z}_1\nonumber\\
\widehat{W}_{obs-hid} &=& \lambda \widehat{z}_1 \widehat{N}^c \widehat{N}^c/2
\end{eqnarray}
where, like $M_N$, $\lambda$ is a matrix in the space of
right-handed neutrino flavors. This superpotential respects a
global U(1)$_{{\rm R-sym}}\otimes$U(1)$_{{\rm Lepton}}$ invariance
in addition to  the baryon number conservation and the MSSM gauge
symmetries.

The supergravity Lagrangian is based on the K{\"a}hler potential \cite{sugra}
\begin{eqnarray}
{\cal{G}}(\widehat{\phi}, \widehat{\phi}^{\dagger}) = K(\widehat{\phi},\widehat{\phi}^{\dagger}) + M_{Pl}^2 \ln
\left|\frac{\widehat{W}(\widehat{\phi})}{M_{Pl}^3}\right|^2
\end{eqnarray}
where $\widehat{\phi}=\left(\widehat{z}_i; \widehat{H}_u,
\widehat{H}_d, \cdots, \widehat{N}^c\right)$ collectively denotes
the chiral superfields in the hidden and observable sectors of the
theory. The kinetic terms of the superfields are collected in
$K(\widehat{\phi},\widehat{\phi}^{\dagger})$. The part of the
scalar potential induced by $F$-terms is given by
\begin{eqnarray}
\label{pot} V_{F}(\phi)=e^{K/M_{Pl}^2} \left\{ g^{a b^{\star}}
D_aW D_{b^\star}W^{\star} - 3 \frac{\left|
W\right|^2}{M_{Pl}^2}\right\}
\end{eqnarray}
where
\begin{eqnarray}
D_a W&=& \frac{\partial W}{\partial \phi^a} + \frac{W}{M_{Pl}^2}
\frac{\partial K}{\partial \phi^a}
\end{eqnarray}
and $g_{a b^{\star}}$ is the K{\"a}hler metric:
\begin{equation}
g_{a b^{\star}}=\frac{\partial^2 K}{\partial\phi^a
\partial\phi^{\star\, b}}.
\end{equation}
 In addition to
$V_F$, there are contributions from $D$-terms as well. However,
$D$-terms do not contribute to supersymmetry breaking since, in
our model, none of the hidden sector superfields exhibits a gauge
invariance.

Similarly to $\widehat{W}$, we decompose kinetic terms of the
superfields as
\begin{eqnarray}
\label{kahlers} K_{obs} &=& \widehat{H}_u^{\dagger} \widehat{H}_u + \widehat{H}_d^{\dagger}
\widehat{H}_d + \cdots + \widehat{N}^{c\, \dagger} \widehat{N}^c\nonumber\\
K_{hid} &=& \widehat{z}_1^{\dagger} \widehat{z}_1 + \widehat{z}_2^{\dagger} \widehat{z}_2 + C_1 M_{Pl}^2
\log\frac{\widehat{z}_1^{\dagger} \widehat{z}_1}{M_{Pl}^2} + C_2 M_{Pl}^2 \log\frac{\widehat{z}_2^{\dagger}
\widehat{z}_2}{M_{Pl}^2} \nonumber\\
K_{obs-hid} &=& \frac{1}{M_{Pl}^2}\left(\lambda_1 \widehat{z}_1^{\dagger} \widehat{z}_1 + \lambda_2
\widehat{z}_2^{\dagger} \widehat{z}_2\right) \widehat{H}_u\cdot \widehat{H}_d + \mbox{h.c.}
\end{eqnarray}
where $K_{obs-hid}$ is similar to the operator used in the Giudice-Masiero mechanism \cite{gm} which solves the
naturalness problem associated with the $\mu$ parameter\footnote{In
principle, the K\"ahler potential can include higher order terms such as
\begin{eqnarray}
\Delta K_{obs-hid}  = \sum_{m,n>1} {\beta_{mn} \over (m ! n
!)^2}\left({
\hat{z}_1^\dagger \hat{z}_1\over M_{Pl}^2} \right)^m
\left({\hat{z}_2^\dagger \hat{z}_2 \over M_{Pl}^2}\right)^n
\widehat{H}_u\cdot\widehat{H}_d\nonumber
\end{eqnarray}
where $\beta_{m n}$ are dimensionless constants. In our analysis,
we restrict  ourselves to the minimal case and neglect such higher
order effects noticing that they do not alter the scale of
observable sector parameters $e.g.$ the $\mu$ parameter and Higgs
bilinear term to be derived in this section. }. The logarithmic terms in
$K_{hid}$, which might be inspired from strings \cite{string}, do
not change the K{\"a}hler metric and are included to achieve a
sensible vacuum configuration in the hidden sector\footnote{ In
general, quantum gravitational interactions do not respect global
symmetries \cite{quantum}, and thus, K{\"a}hler potential above
can receive corrections of the form
\begin{eqnarray}
\Delta K_{hid} = \sum_{m,n,p,q>2}\frac{\alpha_{m n p q}}{M_{Pl}^{m+n+p+q-2}} {(\hat{z}_1^\dagger)^m \hat{z}_1^n
(\hat{z}_2^\dagger)^p \hat{z}_2^q \over m ! n !p!q!}\nonumber
\end{eqnarray}
where $\alpha_{m n p q}$ are dimensionless constants.  Recalling the fact
that global symmetry breaking
effects
get strongly suppressed if gravity is modified near the Planckian
scale
 \cite{reneta}, throughout this work
we neglect such terms. }. The dimensionless
couplings $C_1$ and $C_2$ are determined from the minimization of (\ref{pot})
and demanding zero (or very small) cosmological constant.

 The scalar potential of the hidden sector fields (\ref{pot}) takes the form
\begin{eqnarray}\label{Vf}
V_F({z}_1, {z}_2)&=& M_{hid}^2 \exp\left[\frac{|{{z}}_1|^2+|{{z}}_2|^2}{M_{Pl}^2}\right] \left[
\frac{|{{z}}_1|^2}{M_{Pl}^2} \right]^{C_1} \left[
\frac{|{{z}}_2|^2} {M_{Pl}^2} \right]^{C_2}\nonumber\\
&\times& \left[ |{{z}}_1|^2\left(1+C_2+\frac{|{{z}}_2|^2}{M_{Pl}^2}\right)^2
+|{{z}}_2|^2\left(1+C_1+\frac{|{{z}}_1|^2}{M_{Pl}^2}\right)^2 -3\frac{|{{z}}_1|^2 |{{z}}_2|^2 }{M_{Pl}^2}\right]
\end{eqnarray}
where $z_{1,2}$ stand for the scalar components of $\widehat{z}_{1,2}$, respectively. (In what follows, we will
denote their fermionic partners by $\psi_{z_{1,2}}$.) Clearly, $V_F$ diverges as $|{z}_1|, |{z}_2| \rightarrow
\infty$. Moreover, when $C_1, C_2 < 0$ and $C_1 + C_2 < -1$ potential is not minimized for vanishing ${z}_1$ and
${z}_2$. For determining the vacuum configuration, we should solve
\begin{eqnarray}
\frac{\partial V_F}{\partial |{z}_1|^2} = 0\;, \; \; \frac{\partial V_F}{\partial |{z}_2|^2} = 0\;,\;\;
V_F({z}_1,{z}_2) = 0
\end{eqnarray}
where the first two determine the extremum of the potential whereas the third  is needed for nullifying the
cosmological constant. These conditions lead to the constraints
\begin{eqnarray}\label{1+B+1+A}
1+C_2=\frac{|\langle {{z}}_2\rangle |^2}{M_{Pl}^2} \;,\; 1+C_1= \frac{|\langle {{z}}_1\rangle |^2}{M_{Pl}^2}\;,\;
|\langle {{z}}_1 \rangle|^2+|\langle {{z}}_2 \rangle|^2=\frac{3}{4} M_{Pl}^2
\end{eqnarray}
where  $F$ components of $\widehat{z}_{1,2}$ also develop VEVs
\begin{eqnarray}
\label{fterms} \langle F_{z_1}\rangle &=&2 M_{hid}^* \langle {z}_2^*\rangle \frac{|\langle {z}_1 \rangle|^2}
{M_{Pl}^2}
\exp\left[\frac{K}{2 M_{Pl}^2}\right] \sim M_{Pl}M_{hid}\nonumber\\
\langle F_{z_2} \rangle &=& 2 M_{hid}^* \langle {z}_1^*\rangle \frac{|\langle {z}_2 \rangle|^2} {M_{Pl}^2}
\exp\left[\frac{K}{2 M_{Pl}^2}\right]\sim M_{Pl}M_{hid}\ .
\end{eqnarray}

The vacuum configuration is symmetric under simultaneous $(C_1
\leftrightarrow C_2)$ and $(\left|\langle {z}_1^2 \rangle\right|
\leftrightarrow \left|\langle {z}_2^2 \rangle\right|)$ exchanges.
Clearly, $C_1$ and $C_2$ have to add up to $-5/4$. The vanishing
of the cosmological constant puts stringent constraints on the
allowed ranges of $C_1$ and $C_2$. Indeed, in order to have a
nontrivial vacuum with $\langle {z}_{1,2}\rangle, \langle
F_{z_1,z_2} \rangle \ne 0$ and with vanishing energy, one needs
$-1 < C_2 < -1/4$. (Notice that if we set $V_F$ nonzero but equal
to an exceedingly small value corresponding to the observed
cosmological constant, $C_1$, $C_2$ and $\langle z_{1,2}\rangle $
get modified only slightly, leaving the overall argument similar
to the case  $V_F=0$.) As an explicit example, let us consider the
case $(C_1,C_2)=(-5/8, -5/8)$ which gives rise to $\left|\langle
{z}_1^2 \rangle\right| = \left|\langle {z}_2^2 \rangle\right| =
(3/8) M_{Pl}^2$ at which $V({z}_1,{z}_2) = 0$ as desired, and the
matrix $M_{i,j} =
\partial^2 V_{F}({z}_1,{z}_2)/\partial |{z}_i|^2
\partial |{z}_j|^2$ acquires positive eigenvalues $(5.41, 5.41)$,
 guaranteeing thus the minimization of the potential. (Obviously, there is
nothing special about these numerical values we assign to $C_1$
and $C_2$; they are picked up just for a fast analysis of the
potential landscape.)  Another example is $(C_1,C_2)=(-75/76,
-5/19)$  which yields $\left|\langle {z}_1^2 \rangle\right| =
0.013 M_{Pl}^2$ and $\left|\langle {z}_2^2 \rangle\right| = 0.737
M_{Pl}^2$ at which $V({z}_1,{z}_2) = 0$ as desired, and $M_{i j}$
develops positive eigenvalues $(242.71, 4.34)$.  These case
studies illustrate the behavior of the potential landscape as a
function of $C_1$ and $C_2$.

We can rewrite the $F$-terms in
(\ref{fterms}) as
\begin{eqnarray}
\label{Fterms2} \langle F_{z_1}\rangle &=& 2 e^{3/8}
\left(1+C_1\right)^{1+ \frac{1}{2} C_1} (1+C_2)^{\frac{1}{2} C_2}
\times \langle {z}_2^{\star} \rangle\times M_{hid}^* \nonumber\\
\langle F_{z_2}\rangle &=& 2 e^{3/8} (1+C_1)^{\frac{1}{2} C_1} (1+C_2)^{1+ \frac{1}{2} C_2} \times \langle
{z}_1^{\star} \rangle \times M_{hid}^*
\end{eqnarray}
which are ${\cal{O}}\left(M_{hid} M_{Pl}\right)$. Indeed, for $(C_1,C_2)=(-5/8, -5/8)$ it turns out that
$\left|\langle F_{z_1}\rangle \right| =  \left|\langle F_{z_2}\rangle \right| \approx 1.25 M_{hid} M_{Pl}$, and
for $(C_1,C_2)=(-75/76, -5/19)$ it is that $\left|\langle F_{z_1}\rangle \right| \approx 0.30 M_{hid} M_{Pl}$ and
$\left|\langle F_{z_2}\rangle \right| \approx 2.16 M_{hid} M_{Pl}$. In the latter, $\left|\langle F_{z_1}\rangle
\right|$ is smaller than $\left|\langle F_{z_2}\rangle \right|$ by an order of magnitude due to relative
smallness of $\left|\langle {z}_1^2 \rangle\right|$. The gravitino mass obeys the relation
\begin{eqnarray}
m_{3/2}^2=M_{Pl}^2 e^{{\cal{G}}/M_{Pl}^2} = e^{3/4} (1+C_1)^{1+C_1} (1+C_2)^{1+C_2} M_{hid}^2
\end{eqnarray}
which yields $m_{3/2} \sim M_{hid}$. Indeed, it gives $m_{3/2} =
M_{hid}$ for $(C_1,C_2)=(-5/8, -5/8)$, and $m_{3/2} \approx 1.26
M_{hid}$ for $(C_1,C_2)=(-75/76, -5/19)$. Consequently, the mass
parameter $M_{hid}$ in $\widehat{W}_{hid}$ corresponds to  the
gravitino mass. In other words, $M_{hid} \simeq m_{3/2}$ in the
superpotential and the fundamental scale of gravity $M_{Pl}$
combine to break supersymmetry at the intermediate scale
$M_{\tinySUSY}^2 \sim M_{hid} M_{Pl}$ as suggested by the sizes of
the associated $F$-terms (\ref{Fterms2}).

The spontaneous breakdown of local supersymmetry induces soft
breakdown of global supersymmetry in the observable sector
(presumably with additional hidden sector fields different from
$\widehat{z}_i$ which facilitate induction of $\mu$ and $M_N$ in
the present model) such that each of the scalar fields acquires a
mass-squared $\sim m_{3/2}^2$ and each Yukawa interaction in
$\widehat{W}_{obs}$ gives rise to a triscalar coupling $\sim
m_{3/2}$ \cite{sugra}.

Concerning the parameters pertaining to the right-handed neutrino
sector, from Eq. (\ref{yukawa}), one finds that the right-handed neutrinos
 acquire a mass term in the superpotential
\begin{eqnarray}
\frac{M_N }{2} \widehat{N^c}\widehat{N^c}
\ \ \ \  {\rm where} \ \ \ \
M_{N} = \lambda^{\prime} \langle {z}_1 \rangle
\end{eqnarray}
 and
\begin{eqnarray}
\label{lambdaprime} \lambda^{\prime} \equiv e^{\frac{K}{2
M_{Pl}^2}} \lambda = e^{3/8} (1+C_1)^{C_1/2} (1+C_2)^{C_2/2} \lambda
\end{eqnarray}
in accord with the fact that, in supergravity framework,
observable Yukawa couplings are related to the bare ones in the
superpotential via K{\"a}hler dressing. Moreover, scalar
right-handed neutrinos acquire the bilinear coupling
\begin{eqnarray}
\label{b-term}
\frac{B_N}{2} \widetilde{N}^c \widetilde{N}^c
\ \ \ \ {\rm where} \ \ \ \  B_N = \lambda^{\prime} \langle F_{z_1} \rangle.
\end{eqnarray}
 In terms
of $\lambda^{\prime}$, $M_{N} \approx 0.61 \lambda^{\prime}
M_{Pl}$ and $B_{N} \approx 1.25 \lambda^{\prime} M_{hid} M_{Pl}$
for $(C_1,C_2)=(-5/8, -5/8)$, and $M_{N} \approx 0.11
\lambda^{\prime} M_{Pl}$ and $B_{N} \approx 0.30 \lambda^{\prime}
M_{hid} M_{Pl}$ for $(C_1,C_2)=(-75/76, -5/19)$. The model, in
general, predicts
\begin{eqnarray}
B_N M_N^{-1} = 2 m_{3/2}
\end{eqnarray}
from which it follows that $B_N$ falls in a range that is too
large for ``soft leptogenesis" to be effective \cite{Yuval}. On the
other hand, $B_N$ is too small for inducing significant radiative
effects \cite{farzan}. These numerical estimates are intended for
consistency check of the model.

The seesaw-induced neutrino masses are given by \cite{seesaw}
\begin{eqnarray}
m_\nu = {Y^{\prime}}_\nu^T M_{N}^{-1} {Y^{\prime}}_\nu \langle H_u\rangle^2.
\end{eqnarray}
where again ${Y^{\prime}}_\nu=e^{\frac{K}{2 M_{Pl}^2}} Y_\nu$. One finds that in order to have
 $m_\nu\simeq
0.05\, {\rm eV}$, for $\left|Y^{\prime}_{\nu}\right| {\simlt} 1$,
the right-handed neutrino masses
 must satisfy $M_N \simlt
10^{15}\, {\rm GeV}$. This implies that $(0.1 - 1) \lambda^{\prime} \simlt 10^{-3}$ or $\lambda^{\prime}
\simlt
10^{-2} (0.1 - 1)$. For the neutrino Yukawa matrix $Y^{\prime}_{\nu}\sim{\cal{O}}(1)$,  $M_N$ reaches
$10^{15}\, {\rm GeV}$ level and hence $\lambda^{\prime}$ takes its maximal value.

With the minimal form of the K{\"a}hler potential given in (\ref{kahlers}), the effective $\mu$ parameter of the
MSSM is generated to be
\begin{eqnarray} \mu &=& \frac{1}{M_{Pl}^2} \left( \lambda_1 \langle {z}_1 \rangle
\langle F_{z_1}^* \rangle + \lambda_2 \langle {z}_2 \rangle \langle F_{z_2}^* \rangle\right)\cr &=&2{ M_{hid}
\langle {z}_1\rangle \langle {z}_2 \rangle \over M_{Pl}^4} e^{K/2M_{Pl}^2} \left( \lambda_1 |\langle {z}_1
\rangle|^2+\lambda_2 |\langle {z}_2 \rangle|^2\right) \simeq M_{hid}\label{mumu}
\end{eqnarray}
which lies at the desired scale. This solution for naturalness of
the scale of the $\mu$ parameter\footnote{It might be instructive
to contrast the model advocated here with low-scale MSSM
extensions (by which we mean ($i$) the NMSSM which extends the
MSSM with a chiral singlet superfield by forbidding a bare $\mu$
in the superpotential with a $Z_3$ symmetry, and ($ii$) the
U(1)$^{\prime}$ models which extend the MSSM by both a chiral
singlet and an additional Abelian invariance) which also solve the
$\mu$ problem in a dynamical fashion. The NMSSM suffers from the
cosmological domain wall problem, and U(1)$^{\prime}$ models spoil
gauge coupling unification unless one introduces a number of
exotic or family non-universal charge assignments. The present
model is devoid of such problems as it utilizes global continuous
symmetries operating at high scale.} is similar to the one
proposed in \cite{gm}. The corresponding soft supersymmetry
breaking Higgs bilinear mass-squared parameter reads as
\begin{eqnarray}
B_H &=& \frac{1}{M_{Pl}^2} \left( \lambda_1 \left|\langle F_{z_1} \rangle\right|^2 + \lambda_2 \left|\langle
F_{z_2} \rangle\right|^2\right)\cr &=& 4{\left| M_{hid} \langle {z}_1\rangle \langle {z}_2 \rangle\rangle
\right|^2 \over M_{Pl}^6} e^{K/M_{Pl}^2} \left( \lambda_1 |\langle {z}_1 \rangle|^2+\lambda_2 |\langle {z}_2
\rangle|^2\right) \simeq M_{hid}^2 \label{bh}
\end{eqnarray}
 which is again at the right scale for keeping the MSSM Higgs
sector sufficiently light.

The main implication of the model at hand is that $\mu$ parameter
of the MSSM  and right-handed neutrino masses are correlated with
each other. It might be instructive to illustrate this correlation
explicitly, and this is done in Fig. 1 by plotting $\mu$ and
$m_{3/2}$  as functions of $M_{N}$ for given values of
$\lambda^{\prime}$, $\lambda_1$ and $\lambda_2$. We take
$\lambda^{\prime} \sim 10^{-4}$ so that for any value of $M_N$,
$\langle z_1 \rangle [= (\lambda^{\prime})^{-1} M_{N}]$ and in
turn $\langle z_2 \rangle$ [see Eq. (\ref{1+B+1+A})] can be
calculated . Then, as suggested by the figure, for $M_{hid}= 0.5\
{\rm TeV}$ the $\mu$ parameter exhibits a strong dependence on
$M_N$ depending on the values of $\lambda_1$ and $\lambda_2$. Indeed,
as can be confirmed by using (\ref{mumu}), the $\mu$ parameter
remains around a ${\rm TeV}$ for $\lambda_1=\lambda_2 =1$ whereas
it swings between the unphysical value zero  and the desired value
${\rm TeV}$ when either $\lambda_1$ or $\lambda_2$ vanishes. The
reason is that $\mu$ vanishes at zero $\langle z_1 \rangle$.
In conclusion, the model offers a manifest correlation between
$\mu$ and $M_N$, and $\mu$ gets properly stabilized to lie at a
${\rm TeV}$ when both $\lambda_1$ and $\lambda_2$ are
nonvanishing.
\begin{figure}
\psfig{figure=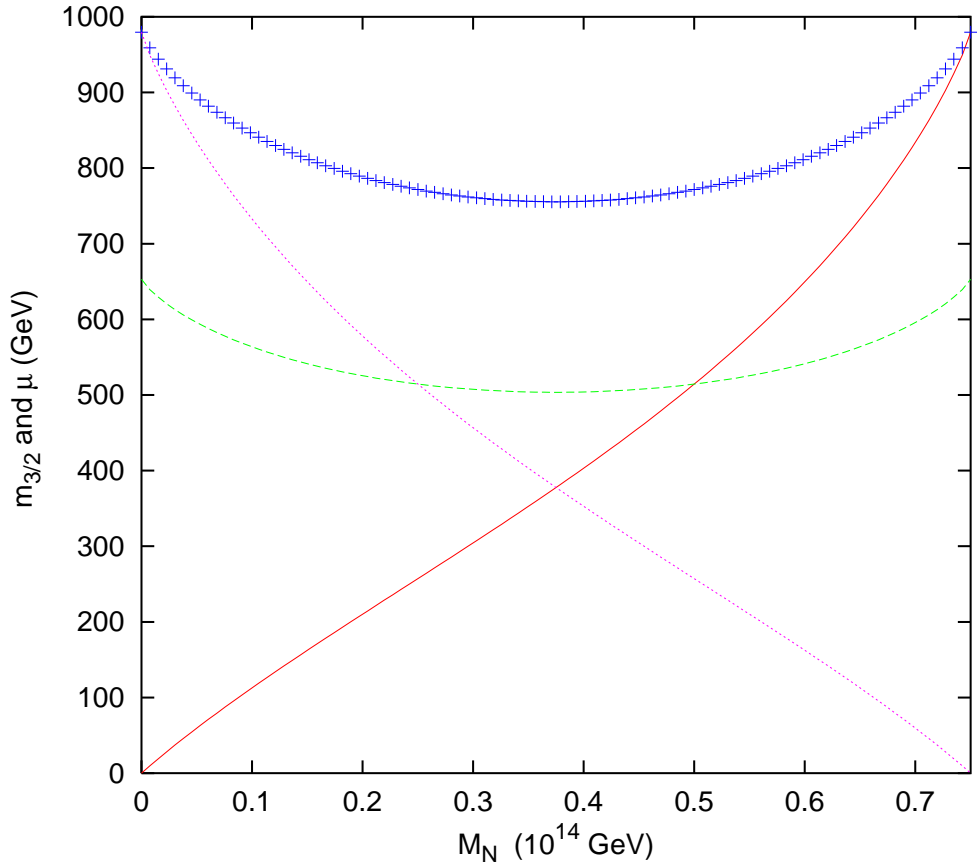,bb=56 35 560 506, clip=true,
 height=5 in
 }
\caption{Variations of $\mu$ and $m_{3/2}$ with $M_N$ for
different values of the model parameters. We have set
  $\lambda^{\prime}={10^{14}\ {\rm GeV} \over M_{Pl}}$ and $M_{hid}=500$ GeV.
The solid and dotted lines show $\mu$ for ($\lambda_1=1,
\lambda_2=0$) and ($\lambda_1=0, \lambda_2=1$), respectively. The
curve depicted with crosses shows $\mu$ for
($\lambda_1=\lambda_2=1$). The dashed line stands for $m_{3/2}$.
  }\end{figure}

For a detailed analysis of constraints on the model discussed in this section, it is necessary to confront it
with the available laboratory, astrophysical and cosmological experimental data. In the next section we will
provide a brief tour of the implications of the present model  for a number of observables.

\section{Phenomenological Implications}
In this section we briefly discuss some phenomenologically
interesting aspects of the model. The following remarks are in
order:
\begin{itemize}

\item  First of all, nonzero $\langle {z}_1\rangle$ and $\langle {z}_2\rangle$
lead to a spontaneous breakdown
of the lepton number conservation and $R$ invariance in the hidden sector. The spontaneous lepton number breaking releases a
massless Goldstone boson, the Majoron $J$:
\begin{eqnarray}
\label{z1} {z}_1 =(\mid \langle {z}_1 \rangle\mid + \phi_1)\times \exp{i \left(\mbox{Arg}\left[\langle {z}_1
\rangle\right] + \frac{J}{\mid \langle {z}_1 \rangle\mid}\right)}
\end{eqnarray}
where
the  real scalar fields $\phi_1$ and $J$ denote fluctuations about
the vacuum state. In the flavor basis, the Majoron couples to
right-handed neutrinos via $\lambda J N^T C N$; however, it does
not couple to the left-handed active neutrinos. This means that in
the mass basis, the light active neutrinos couple to the Majoron
with an exceedingly small strength $\sim \lambda Y_{\nu}^2 \langle
H_u \rangle^2/M_{N}^2$. This coupling is too small to have any
significance \cite{majoron}.

In this model, due to the spontaneous breakdown of $R$-symmetry there is an additional Goldstone boson, $J'$,
\begin{eqnarray}
\label{z2} {z}_2 =(\mid \langle {z}_2 \rangle\mid + \phi_2)\times \exp{i \left(\mbox{Arg}\left[\langle {z}_2
\rangle\right] + \frac{J^{\prime}}{\mid \langle {z}_2 \rangle\mid}\right)}
\end{eqnarray}
where we have used  a parametrization similar to (\ref{z1}). The coupling of
$J^{\prime}$  to the SM particles is determined by the K\"ahler
potential and is suppressed by $m_{SUSY}/M_{Pl} \sim
10^{-15}-10^{-16}$ which is again too small to have any
phenomenological consequences.

In the supersymmetric Majoron model infamous smajoron problem
arises. In the following, we contrast the present model with
supersymmetric Majoron model and compare status of smajoron
problem in two scenarios. In the singlet Majoron model
\cite{majoron2} there exists a new superfield $\widehat{S}$ which
carries 2 units of lepton number and  couples to the right-handed
neutrinos via $W_1= \hat{S} \hat{N^c} \hat{N^c}$. In similarity
with the model proposed in this work, the right-handed neutrinos
acquire masses through the VEV of $\widetilde{S}$. However, in
this singlet Majoron model the mechanism responsible for
nonvanishing VEVs is different from the one  in our model: To develop VEVs
additional superfields $\widehat{S}^{\prime}$ and
$\widehat{\Lambda}$, with respective lepton numbers equal to -2
and 0, are introduced such that they interact through the
superpotential $W_2= \widehat{\Lambda}
(\widehat{S}\widehat{S}^{\prime}- M^2).$ It can be shown that
$\langle \widetilde{\Lambda} \rangle=0$ while $\langle
\widetilde{S} \rangle, \langle \widetilde{S}^{\prime} \rangle \neq
0$. Since $\langle \widetilde{\Lambda}\rangle=0$, one out of three
linear combinations of the fermionic components of
$\widehat{\Lambda}$, $\widehat{S}$ and $\widehat{S}^{\prime}$ is
massless, and  can be interpreted as Goldstino. However, in the
present model, the mass matrix of $\psi_{z_1}$ and $\psi_{z_2} $
(the fermionic components of $\widehat{z}_1$ and $\widehat{z}_2$) is
equal to
\begin{eqnarray}
e^{\frac{K}{2 M_{Pl}^2}} \left( \matrix{0 & M_{hid} \cr M_{hid} & 0} \right)
\end{eqnarray}
and  does not possess any vanishing eigenvalue. Namely, massless Majoron does not have any fermionic
counterpart. This difference between the two models is not surprising at all since
in the singlet Majoron model,
supersymmetry is preserved \cite{majoron2} ($\langle F_{S}\rangle=\langle F_{S'}\rangle=\langle
F_{\Lambda}\rangle=0$) and a massless bosonic particle has to have a
fermionic partner whereas in our
model supersymmetry is broken ($\langle F_{z_1}\rangle,\langle F_{z_2} \rangle \ne 0$).

\item Notice that spontaneous breakdowns of lepton number conservation and $R$
invariance in the hidden sector reflect themselves as explicit
breaking in the observable sector. Indeed, as is clear from
$\widehat{W}_{obs-hid}$, $\langle z_1 \rangle \neq 0$ leads to
explicit lepton number breaking via the induced right-handed
neutrino mass. On the other hand, $\langle F_{z_{1,2}}\rangle \neq
0$ lead to spontaneous breakdown of supergravity whereby inducing
explicit soft-breaking terms \cite{dimopoulos}. These soft terms
lead to explicit breaking of $R$ invariance down to its $Z_2$
subgroup, the $R$-parity, which is an exact discrete symmetry of
the observable sector. Indeed, $\langle F_{z_1} \rangle$, for
instance, induces a neutrino B-term which explicitly breaks the
$R$ invariance but conserves the $R$ parity. As a result, the
model accommodates a natural candidate for cold dark matter  which
is the famous lightest supersymmetric particle.

\item  In the present model, $z_i$ and $\psi_{z_i}$ ($i=1,2$) acquire masses of order of $m_{3/2}$.
On the other hand, their couplings to MSSM spectrum are rather weak. Therefore, their decay rates are expected to
be suppressed. For instance, their decay into Higgs fields occur with a rate
\begin{eqnarray}
\Gamma (z_i \to H  H)\sim {M_{hid} \over 4 \pi}
 \left( {M_{hid} \over M_{Pl}}\right)^2
\end{eqnarray}
which is suppressed by the small ratio $\left( {M_{hid} \over M_{Pl}}\right)^2\sim 10^{-32}$. The smallness of
$\widehat{z}_i$ decay rates into SM species is rather generic.

At first sight, it may seem that this low but nonzero decay rate
may be problematic  for the  big bang nucleosynthesis, especially
in the face of the fact that at temperatures higher than $M_N$
these particles can be produced via $NN\to z z$ at a rate  not
suppressed by $M_{Pl}$. Fortunately, a simple  estimate shows that
initial abundance of these particles cannot take too high values:
\begin{eqnarray}
{m_z n_z \over n_\gamma} \sim m_z\Gamma (N \to zz)\,
t\mid_{T=M_N} &\sim& \left({|\lambda^{\prime}|^4\over 4 \pi^3}
T\right)\left( 0.2 {M_{Pl} \over {\rm
MeV}^2}(\frac {{\rm MeV}}{T})^2\right)m_z\nonumber\\
&\sim& \left({m_z \over 1000\ {\rm  GeV}}\right) \left( {\lambda'
\over 10^{-4}} \right)^3 \times 10^{-12} \ {\rm GeV}\,.
\end{eqnarray}
According to Fig. 3 of \cite{lopez}, the late decay of such
particles cannot destroy the results of big bang nucleosynthesis.

\item  It might be instructive to analyze under what conditions one can make $\langle {z}_1\rangle$ and
$\langle {z}_2\rangle$ hierarchically split so that $M_N \sim 10^{15}\, {\rm GeV}$ arises with
$\lambda^{\prime} \sim {\cal{O}}(1)$. This can be achieved
with a sufficiently small $1+C_1$. Setting $C_1 = -1 + \epsilon^2$ with $\epsilon^2 \ll 1$ , Eq.
(\ref{1+B+1+A}) implies
\begin{eqnarray}
\label{epsilon-sonuc} && \langle {z}_1 \rangle \sim  \epsilon M_{Pl} \;,\;\; \langle F_{z_1} \rangle
\sim \epsilon
M_{Pl}M_{hid}\nonumber\\
&& \langle {z}_2 \rangle \sim   M_{Pl} \;,\;\; \langle F_{z_2} \rangle \sim  M_{Pl}M_{hid}
\end{eqnarray}
which make it manifest that, for a sufficiently
small $\epsilon$,  $\langle {z}_1 \rangle$ and $\langle F_{z_1} \rangle $ can be substantially
smaller than, respectively,  $\langle {z}_2 \rangle$ and $\langle F_{z_2} \rangle $.  In fact, the VEVs in (\ref{epsilon-sonuc}) suggest that
\begin{eqnarray}
M_N = \lambda^{\prime} \langle {z}_1 \rangle \sim \epsilon \lambda^{\prime} M_{Pl}\;,\;\; B_N=\lambda^{\prime}
\langle F_{z_1} \rangle \sim \lambda^{\prime} \epsilon M_{Pl} M_{hid}
\end{eqnarray}
both of which involve $\epsilon M_{Pl}$ rather than $M_{Pl}$ itself. This $\epsilon$ dependence, however, is not
present in the Higgs sector
\begin{eqnarray}
\label{higgssector} \mu \simeq \lambda_2 \frac{\langle{z}_2\rangle \langle F_{z_2}\rangle}{ M_{Pl}^2}\sim
 M_{hid} \;,\;\; B_H \simeq \lambda_2 \frac{|\langle F_{z_2} \rangle|^2}{M_{Pl}^2}\sim
M_{hid}^2
\end{eqnarray}
and $m_{3/2} \sim M_{hid}$ still holds.
Finally, one notes that ${z}_{i}$ and their fermionic
counterparts weigh now $M_{hid}/\epsilon$ rather than $M_{hid}$.
Taking  $\epsilon \sim
10^{-3}$, the existing neutrino data can be explained with $\lambda^{\prime} \sim 1$.

\item Let us analyze the Higgs sector in more detail. Note that, without loss of generality, we
can absorb the phase of $M_{hid}$ by rephasing $\widehat{z_2}$.
Therefore, hereon we are going to assume that $M_{hid}$ is real.
First consider the phase of the $\mu$ parameter. From (\ref{bh})
and (\ref{mumu}) it follows that the relative phase between $\mu$
and $B_H$ is determined solely by the total phase of $\langle
z_1\rangle \langle z_2 \rangle$. More explicitly,
\begin{eqnarray}
\label{phase} \mbox{Arg}\left[\frac{\mu}{B_H}\right] = \mbox{Arg}\left[\langle z_1 \rangle\right] +
\mbox{Arg}\left[\langle z_2 \rangle\right]
\end{eqnarray}
as follows from (\ref{z1}) and (\ref{z2}). This is a physical
basis-independent phase that cannot be rotated away by
redefinition of the relative phase between $\widehat{H}_u$ and $\widehat{H}_d$.
In the context of  both constrained \cite{cmssm} and unconstrained \cite{umssm} MSSM, the present bounds
on electric
dipole moments imply that  $\mbox{Arg}\left[\langle z_1 \rangle  \langle z_2 \rangle\right]$ cannot
exceed  a
few percent. Therefore, the VEVs $\langle
z_1 \rangle$ and $\langle z_2 \rangle$ must be nearly back-to-back in the vacuum of the theory.

The structure of the model entails certain correlations among certain parameters which would bear no correlation
in the MSSM, constrained or otherwise. An interesting case concerns the phase of the $\mu$ parameter. To see this,
consider the Dirac coupling of sneutrinos in the soft-breaking sector:
\begin{eqnarray}
{\cal{L}} \ni a_0 Y_{\nu}^{\prime} \widetilde{L}\cdot H_u \widetilde{N^c} + \mbox{h.c.}
\end{eqnarray}
which respects both the lepton number and $R$-parity. Here $a_0$ is the universal trilinear coupling, as appropriate
for the constrained MSSM. It can be shown that at two-loop level this operator induces a bilinear interaction
between $z_1$ and $z_2$ as follows
\begin{eqnarray}
\label{bz} {\cal{L}} \ni c_z^{\star} M_{hid} z_1 z_2 + \mbox{h.c.}
\end{eqnarray}
where
\begin{eqnarray}
\label{cz*} c_z^{\star}=(\mbox{an} \; {\cal{O}}(1)\; \mbox{real}\;
 \mbox{number})\times a_0 {\lambda^{\prime\, T}
Y_\nu^{\prime\,\star} Y_\nu^{\prime\, T}\lambda^{\prime\, \star} \over (16 \pi^2)^2}
\end{eqnarray}
which is a small perturbation in size. However, an interaction of
the form (\ref{bz}) leads to direct alignment of $\langle z_1
\rangle \langle z_2 \rangle$ towards $c_z$ for scalar potential
(in terms of $H_u^0, H_d^0, \widetilde{\nu}_L^i,
\widetilde{N^c}_i$) to possess a local minimum. Therefore, the
phase of $\mu$ ($i.e.$ $\mbox{Arg}\left[\langle z_1 \rangle
\langle z_2 \rangle\right]$) relaxes to that of $c_z$ and hence to
the phase
 of $a_0$ according to  (\ref{cz*}). (Note that the combination
$\lambda^{\prime\, T}Y_\nu^{\prime\,\star} Y_\nu^{\prime\,
T}\lambda^{\prime\, \star}/ (16 \pi^2)^2$ is real.) Consequently,
the phase of $\mu$ gets traded for that of the trilinear
couplings, and as a by-product of this
correlation, in future if a nonzero $d_e$ is measured, within this
model, we can extract the phase of $\mu$ which corresponds to the
phase of $a_0$ and then, in principle, we can predict the values
$d_{Hg}$ and $d_n$ and test these predictions in the laboratory
experiments \cite{yasaman}.

Having discussed the CP--odd phase of the $\mu$ parameter, we now analyze the role of the $B_H$ parameter in
some depth.
In general, $\mu$ parameter can be determined by measuring masses and mixing angles of the charginos and neutralinos.
On the other hand by studying the mass and decay modes of the CP-odd Higgs boson, $A_0$, we can derive the value
of $B_H$ \cite{hunter}. Therefore, the ratio $|B_H/\mu|$ can be measured in a rather model-independent way. It is
straightforward to show that
\begin{eqnarray}
2 M_{hid} <|B_H/\mu|=2m_{3/2}<2.6 M_{hid}
\end{eqnarray}
in the present model. Since interactions of gravitino are suppressed by $M_{Pl}^{-1}$, it cannot, in practice,
 be detected
at colliders. However, gravitino mass affects cosmological observations, opening a window for testing this
relation. Indeed, from the relation
\begin{eqnarray}
|\mu|^2+m_{H_d}^2=B_H \tan \beta -{m_Z^2\over 2} \cos 2 \beta
\end{eqnarray}
we expect $|B_H/\mu|\sim |\mu|/\tan \beta$ which means that at large $\tan \beta$, there is a ``little hierarchy"
among the parameters of the model. Within our model this implies that gravitino,
rather than the
lightest neutralino,
with mass $\sim |\mu|/\tan \beta$, might be the lightest supersymmetric particle and hence a candidate for cold
dark matter  (see, for instance, \cite{olive} for gravitino dark matter in constrained MSSM).
\end{itemize}

The discussions above provide a brief summary of the implications
of the model constructed in Sec. 2. For a proper description of
the phenomenology of this model, it is necessary to perform a
detailed analysis of various quantities of phenomenological
interest.

\section{Conclusion}
In this work we have constructed a hidden sector model, within
$N=1$ supergravity, for generating, upon supergravity breakdown,
the $\mu$ parameter of the MSSM and the right-handed neutrino mass
$M_N$ at their right scales. The model  utilizes global lepton
number conservation and $R$ invariance to forbid bare $\mu$ and
$M_N$ parameters appearing in the superpotential. Moreover, the
model employs a non-minimal K{\"a}hler potential exhibiting
logarithmic dependencies on the hidden sector fields. The origin
of these non-minimal contributions are left unexplained, yet string
compactification has been an inspiring source. We have
determined parameter regions where $M_N$ and $\mu$ come out to lie
at their right scales, and found that the VEVs of the hidden
sector fields can exhibit a hierarchical splitting so as to reduce
unnatural tunings of the superpotential parameters. As footnoted
in the text, the model at hand neither suffers from domain wall
problem,  nor exhibits any tension with gauge coupling
unification as encountered, respectively, in NMSSM and
U(1)$^{\prime}$ models.

We have confronted the model put forward with a number of
observables, and identified distinctive features and ways of
evading the existing bounds from various sources.
The model predicts
$|B_H/\mu|=2m_{3/2}$ which means, for a large part of parameter
space, the   gravitino is the lightest supersymmetric particle and thus
a candidate for dark matter. Consequently, the mechanism advocated
in this work  possesses potential implications for various
observables, a global analysis of which is yet to be
performed.

\section{Acknowledgements}
Authors would like to thank Lotfi Boubekeur, Lisa Everett and
Goran Senjanovi{\v c} for fruitful discussions. We are also
grateful to M.M. Sheikh-Jabbari for careful reading of the
manuscript and useful remarks. D.A.D. is grateful to Institute for
Studies in Theoretical Physics and Mathematics (IPM) for its
generous hospitality while this work was started, and
International Centre for Theoretical Physics (ICTP), Turkish
Academy of Sciences (through GEBIP grant), and Scientific and
Technical Research Council of Turkey (through project 104T503) for
their financial supports. Y. F. would like to thank ICTP for kind
hospitality where part of this work was done.


\begin{thebibliography}{99}
\bibitem{susskind}
L.~Susskind,
  Phys.\ Rev.\ D {\bf 20}, 2619 (1979).

\bibitem{sugra}
S.~K.~Soni and H.~A.~Weldon,
  Phys.\ Lett.\ B {\bf 126}, 215 (1983).

\bibitem{dynamic}
E.~Witten,
  Nucl.\ Phys.\ B {\bf 188}, 513 (1981);
I.~Affleck, M.~Dine and N.~Seiberg,
  Phys.\ Rev.\ Lett.\  {\bf 52}, 1677 (1984).


\bibitem{smirnov}
A.~Yu.~Smirnov,
  arXiv:hep-ph/0411194.


\bibitem{seesaw}
M.~Gell-Mann, P.~Ramond and R.~Slansky,
Print-80-0576 (CERN) (see also: P.~Ramond,
  arXiv:hep-ph/9809459.)
T.~Yanagida,
{\it In Proceedings of the Workshop on the Baryon Number of the Universe and Unified Theories, Tsukuba, Japan,
13-14 Feb 1979};
S.~L.~Glashow,
HUTP-79-A059
{\it Based on lectures given at Cargese Summer Inst., Cargese, France, Jul 9-29, 1979};
 R.~N.~Mohapatra and G.~Senjanovic,
  Phys.\ Rev.\ Lett.\  {\bf 44} (1980) 912.

\bibitem{pq}
R.~D.~Peccei and H.~R.~Quinn,
  Phys.\ Rev.\ Lett.\  {\bf 38}, 1440 (1977);
M.~Dine, W.~Fischler and M.~Srednicki,
  Phys.\ Lett.\ B {\bf 104}, 199 (1981);
A.~R.~Zhitnitsky,
  Sov.\ J.\ Nucl.\ Phys.\  {\bf 31}, 260 (1980)
  [Yad.\ Fiz.\  {\bf 31}, 497 (1980)];
J.~E.~Kim,
  Phys.\ Rev.\ Lett.\  {\bf 43}, 103 (1979);
M.~A.~Shifman, A.~I.~Vainshtein and V.~I.~Zakharov,
  Nucl.\ Phys.\ B {\bf 147}, 385 (1979).

\bibitem{lepto}
M.~Fukugita and T.~Yanagida,
  Phys.\ Lett.\ B {\bf 174}, 45 (1986).

\bibitem{buchmuller}
 W.~Buchmuller, P.~Di Bari and M.~Plumacher,
  Annals Phys.\  {\bf 315}, 305 (2005)
  [arXiv:hep-ph/0401240];
  G.~F.~Giudice, A.~Notari, M.~Raidal, A.~Riotto and A.~Strumia,
MSSM,''
  Nucl.\ Phys.\ B {\bf 685}, 89 (2004)
  [arXiv:hep-ph/0310123]; see however,
Y.~Farzan and J.~W.~F.~Valle,
  arXiv:hep-ph/0509280.

\bibitem{kimnilles}
E.~J.~Chun, J.~E.~Kim and H.~P.~Nilles,
  Nucl.\ Phys.\ B {\bf 370} (1992) 105.


\bibitem{lepsite}
The LEP SUSY working group, http://lepsus.web.cern.ch/lepsusy/.
\bibitem{lower}
G.~Belanger, F.~Boudjema, A.~Cottrant, A.~Pukhov and S.~Rosier-Lees,
  JHEP {\bf 0403}, 012 (2004)
  [arXiv:hep-ph/0310037].




\bibitem{muprob}
P.~Fayet,
Phys.\ Lett.\ B {\bf 69}, 489 (1977).
J.~E.~Kim and H.~P.~Nilles,
Phys.\ Lett.\ B {\bf 138}, 150 (1984);
D.~Suematsu and Y.~Yamagishi,
Int.\ J.\ Mod.\ Phys.\ A {\bf 10}, 4521 (1995) [arXiv:hep-ph/9411239];
M.~Cvetic and P.~Langacker,
Mod.\ Phys.\ Lett.\ A {\bf 11}, 1247 (1996) [arXiv:hep-ph/9602424];
V.~Jain and R.~Shrock,
arXiv:hep-ph/9507238;
Y.~Nir,
Phys.\ Lett.\ B {\bf 354}, 107 (1995) [arXiv:hep-ph/9504312].

\bibitem{dimopoulos}
S.~Dimopoulos and S.~D.~Thomas,
  Nucl.\ Phys.\ B {\bf 465}, 23 (1996)
  [arXiv:hep-ph/9510220];
D.~A.~Demir,
  Phys.\ Rev.\ D {\bf 62}, 075003 (2000)
  [arXiv:hep-ph/9911435].

\bibitem{gm}
G.~F.~Giudice and A.~Masiero,
  Phys.\ Lett.\ B {\bf 206}, 480 (1988).

\bibitem{string}
L.~J.~Dixon, V.~Kaplunovsky and J.~Louis,
  Nucl.\ Phys.\ B {\bf 329}, 27 (1990);
J.~P.~Derendinger, S.~Ferrara, C.~Kounnas and F.~Zwirner,
  Nucl.\ Phys.\ B {\bf 372}, 145     (1992).

\bibitem{quantum}
M.~Kamionkowski and J.~March-Russell,
  Phys.\ Lett.\ B {\bf 282}, 137 (1992)
  [arXiv:hep-th/9202003];
  R.~Kallosh, A.~D.~Linde, D.~A.~Linde and L.~Susskind,
  Phys.\ Rev.\ D {\bf 52}, 912 (1995)
  [arXiv:hep-th/9502069];
 R.~Holman, S.~D.~H.~Hsu, T.~W.~Kephart,
 E.~W.~Kolb, R.~Watkins and L.~M.~Widrow, Phys.\ Lett.\ B {\bf 282}, 132
(1992)  [arXiv:hep-ph/9203206];
S.~M.~Barr and D.~Seckel,
  Phys.\ Rev.\ D {\bf 46}, 539 (1992)
\bibitem{reneta}
R.~Kallosh, A.~D.~Linde, D.~A.~Linde and L.~Susskind,
  Phys.\ Rev.\ D {\bf 52}, 912 (1995)
  [arXiv:hep-th/9502069].
\bibitem{Yuval}
Y.~Grossman, T.~Kashti, Y.~Nir and E.~Roulet,
  Phys.\ Rev.\ Lett.\  {\bf 91}, 251801 (2003)
  [arXiv:hep-ph/0307081];
 G.~D'Ambrosio, G.~F.~Giudice and M.~Raidal,
  Phys.\ Lett.\ B {\bf 575}, 75 (2003)
  [arXiv:hep-ph/0308031].

\bibitem{farzan}
 Y.~Farzan,
  JHEP {\bf 0502}, 025 (2005)
  [arXiv:hep-ph/0411358];
 Y.~Farzan,
  Phys.\ Rev.\ D {\bf 69}, 073009 (2004)
  [arXiv:hep-ph/0310055];
 Y.~Farzan,
  arXiv:hep-ph/0505004.




\bibitem{majoron}
G.~F.~Giudice, A.~Masiero, M.~Pietroni and A.~Riotto,
  Nucl.\ Phys.\ B {\bf 396}, 243 (1993)
  [arXiv:hep-ph/9209296];
 Y.~Farzan,
  Phys.\ Rev.\ D {\bf 67}, 073015 (2003)
  [arXiv:hep-ph/0211375].


\bibitem{majoron2}
R.~N.~Mohapatra and X.~Zhang,
  Phys.\ Rev.\ D {\bf 49}, 1163 (1994)
  [Erratum-ibid.\ D {\bf 49}, 6246 (1994)]
  [arXiv:hep-ph/9307231].



\bibitem{gravitino}
 M.~Bolz, W.~Buchmuller and M.~Plumacher,
  Phys.\ Lett.\ B {\bf 443}, 209 (1998)
  [arXiv:hep-ph/9809381];
M.~Bolz, A.~Brandenburg and W.~Buchmuller,
  Nucl.\ Phys.\ B {\bf 606}, 518 (2001)
  [arXiv:hep-ph/0012052].


\bibitem{bbn}
W.~Buchmuller, R.~D.~Peccei and T.~Yanagida,
  arXiv:hep-ph/0502169.

\bibitem{cmssm}
T.~Ibrahim and P.~Nath,
  Phys.\ Rev.\ D {\bf 58}, 111301 (1998)
  [Erratum-ibid.\ D {\bf 60}, 099902 (1999)]
  [arXiv:hep-ph/9807501];
D.~A.~Demir, O.~Lebedev, K.~A.~Olive, M.~Pospelov and A.~Ritz,
  Nucl.\ Phys.\ B {\bf 680}, 339 (2004)
  [arXiv:hep-ph/0311314].
\bibitem{lopez}
J.~R.~Ellis, G.~B.~Gelmini, J.~L.~Lopez, D.~V.~Nanopoulos and
S.~Sarkar,
  Nucl.\ Phys.\ B {\bf 373}, 399 (1992).
See, however, R.~Allahverdi and A.~Mazumdar,
  arXiv:hep-ph/0512227;
arXiv:hep-ph/0505050.




\bibitem{umssm}
M.~Brhlik, G.~J.~Good and G.~L.~Kane,
  Phys.\ Rev.\ D {\bf 59}, 115004 (1999)
  [arXiv:hep-ph/9810457];
S.~Abel, S.~Khalil and O.~Lebedev,
  Nucl.\ Phys.\ B {\bf 606} (2001) 151
  [arXiv:hep-ph/0103320];



\bibitem{yasaman}
D.~A.~Demir and Y.~Farzan,
  JHEP {\bf 0510}, 068 (2005)
  [arXiv:hep-ph/0508236].

\bibitem{hunter}
J.~F.~Gunion, H.~E.~Haber, G.~L.~Kane and S.~Dawson, ``The Higgs Hunter's Guide,'' SCIPP-89/13, (1989)

\bibitem{olive}
J.~R.~Ellis, K.~A.~Olive, Y.~Santoso and V.~C.~Spanos,
  Phys.\ Lett.\ B {\bf 588}, 7 (2004)
  [arXiv:hep-ph/0312262].



\end{thebibliography}
\end{document}